# Comparative Analysis of Room Temperature Plasmonic Graphene Hot Electron Bolometric Photodetectors

*Jacek Gosciniak[1], and Jacob B. Khurgin[2]*

[1]*Independent Researcher, 90-132 Lodz, Poland*
[2]*John Hopkins University, Baltimore, MD 21218, USA*

**Abstract**
We appraise a waveguide-integrated plasmonic graphene photodetector based on the hot carrier photo-bolometric effect, with performance characterized simultaneously by high responsivity, on the scale of hundreds of A/W, and high speed on the scale of 100's of GHz. Performance evaluation is based on a theory of bolometric effect originating from the band nonparabolicity of graphene. Results compare favorably with the state-of-the-art plasmonic bolometric photodetectors, predicting up to two orders of magnitude increase in a responsivity while keeping speed on the same level, defined by the electron-lattice scattering time in graphene.

**Introduction**
The next generation of datacom and telecom communication systems requires further increasing the bandwidth of the systems while reducing the size, and power consumption [1, 2]. To meet those requirements a new material platform need to me implemented that is integrated with the traditional Si photonics platform [3]. Graphene is a very promising material for signal modulation and photodetection owing to its extraordinary transport properties [4, 5]. Being only one atom thick it absorbs 2.3 % of incident light in a very wide energy spectrum [6, 7]. It has tunable electronic and optical properties [4], fast carrier dynamics [8, 9] and high carrier mobility [10] enabling ultrafast conversion of photons or plasmons to electrical current or voltage [11, 12]. Moreover, graphene is CMOS-compatible [13] allowing integration on wafer-scale [7, 14] to realize, for example, broadband and sensitive photodetectors [15-23] and fast and cost-effective modulators [24-28].
High-speed and high responsivity photodetectors are crucial components in optical communications that convert the absorbed photons into an electrical signal [3, 29]. Being the last components in the optical links, the detectors must operate with low power and beyond 100 GHz. Furthermore, they should be characterized by low dark current and high responsivity. To meet those requirements, it is desired to implement a new material platform with plasmonics [30]. Plasmonics can squeeze light well below a diffraction limit, which reduces the device footprint [31, 32]. Furthermore, the small device volume means a higher density of integration, and simultaneously, lower power consumption, easier heat dissipation, and faster operation speed [20, 31, 33, 34]. During the last few years some waveguide-integrated plasmonic photodetectors on silicon [35-38] and germanium [39, 40] have been proposed and fabricated. However, they suffer either from a low operation speed, low responsivity or large footprint [41]. Thus, in last few years a lot of effort focused on graphene photodetector. They can operate based on photovoltaic [20, 33, 42], photo-thermoelectric [43-47], or photo-bolometric [33, 34, 42, 48-51] effects. The choice of effect depends on a photodetector's configuration and specific applications. Bolometers have emerged as the technology of choice, because they do not need cooling [52]. They suffer, however, from a low operational speed due to their large heat capacitance and high thermal insulation [48]. Low electronic heat capacity of graphene makes it a material of choice for such photodetectors. Most of the previously reported bolometers operate based on the rise of the lattice temperature that is pretty slow what limits the speed of the photodetectors [11, 53]. In contrary, we use here fast rise and decay times of electronic temperature what highly improves the speed of the photodetectors. We utilize the temperature-dependent material properties for photodetection - the incident light raises the local electronic temperature of the material, which reduces the resistance of the device and produces a change in the current [51, 53]. Graphene is well-suited for this purpose as it has a small electron heat capacity and weak electron-phonon (e-ph) coupling leading to a large light-induced change in electron temperature [43, 44, 54]. The low density of states and small volume for a given area result in a large rise of electronic temperature and, in consequence, enhanced responsivity while a reasonably fast e-l scattering time provides fast device response beyond hundreds of GHz [34, 50].

**Bolometric photodetector arrangement**
Most of the previously reported waveguide-integrated photodetectors suffer from weak external responsivities that origin from low input power coupled to the photodetector and the weak light-matter interaction [13-17]. To enhance the light interaction with graphene, the plasmonic structures should be incorporated as they provide strong mode confinement and local field enhancement [20, 24, 30-32]. Furthermore, they scale down the length of the device to a micrometer range. Most of the plasmonic waveguides support the TM mode with a dominant out-of-plane electric field component while the absorption in graphene requires the in-plane electric field component. Thus, finding a proper design that fulfills those requirements is under a deep interest. Furthermore, the realization of bolometric or photo-thermoelectric photodetectors requires an increase of the electronic temperature of graphene [34, 46]. This can be achieved by an efficient conversion of incident power into electronic heat. For compact photodetectors the power delivered to the small area of graphene highly increases the electronic temperature.

In our previous study [34] we developed a theory of the bolometric effect originating from the band nonparabolicity of graphene that does not require establishment of a perfect equilibrium between the electrons and leads to a simple expression for responsivity that depends only on a very few material parameters. Here we apply this theory to some recently proposed photodetector structures with the goal of optimizing performances. We start this analysis with our proposed recently bolometric photodetector [34] that is based on the long-range dielectric loaded surface plasmon polariton (LR-DLSPP) waveguide [35, 39, 43, 55, 56] with graphene placed at the maximum electric field of the propagating mode (Fig. 1). As a result, the SPP energy is mostly absorbed by the graphene sheet causing rise in electronic temperature and change in resistance [57, 58] while the fraction of the SPP lost in the metal is relatively small.

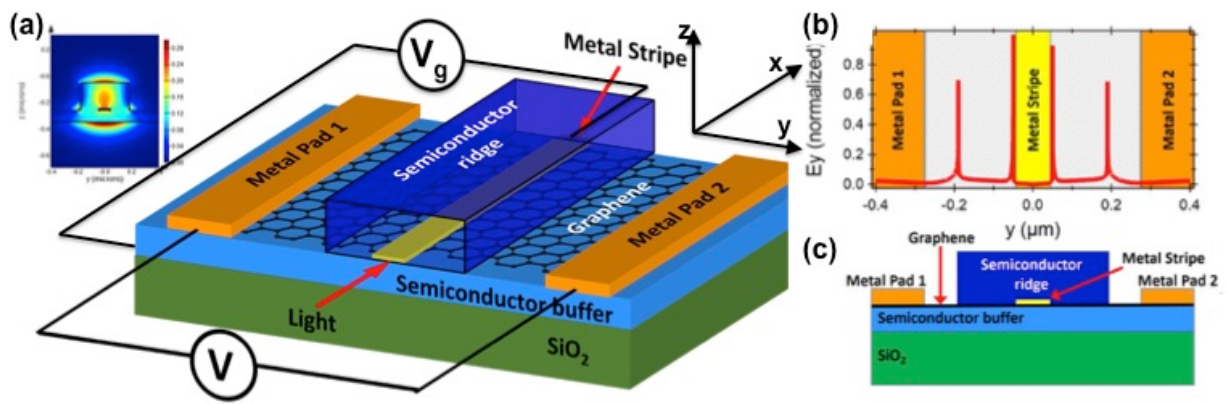

**Fig. 1.** Schematic of the proposed plasmonic bolometric photodetector in symmetric external electrodes arrangement. Here, Si is used as both Semiconductor ridge and Semiconductor buffer material. A distance between Metal Pad 1 and Metal Pad 2 defines the length of the photodetector *L* (along y axis direction), while the length of the waveguide defines the width of the photodetector *W* (along x axis direction). (b) In-plane electric field component of the plasmonic TM mode on graphene sheet and (c) the cross section of the photodetector. Inset in (a) shows the $E^2$ of the TM mode.

As mentioned above, the evaluation is performed based on recently introduced a theory of bolometric effect originating from the band nonparabolicity of graphene [34]. To minimize the contribution from the photo-thermoelectric (PTE) effect, the symmetric contact arrangement has been implemented here, *i.e.*, the same metals were used as contacts and the structure was symmetric with respect to the center of the metal stripe/ridge [Fig. 1b]. As a result, the band diagram across the active graphene channel is symmetric. Consequently, only photo-bolometric (PB) and photo-conductive (PC) effects can exist under this arrangement [34]. The specific of graphene is such that the bolometric effect of increases resistance is always accompanied by the photoconductive effect caused by interband absorption that reduced resistance [34]. However, as it was shown previously [34], the bolometric effect dominates due to the fact that $E_F < \hbar\omega/2$ and $\tau_{ee} < \tau_{el}$.

The in-plane electric field component of the propagating LR-DLSPP mode (Fig. 1b) interacts strongly with graphene enhancing absorption [Supporting Fig. S1]. The presence of the in-plane component of the electric field even for the TM mode is associated here with the small thickness of the metal stripe and its sharp metal corners [24]. The electric field at the metal stripe's corners is very strong but decays quickly on the graphene.

**Evaluation of bolometric photodetectors' performance**

As shown previously [34], the ratio of resistances *ΔR/R*, a key characteristic of graphene bolometric detector performance can be found as

$$\frac{\Delta R}{R} = \frac{\eta_{abs} P_{in} \tau_{el}}{2 E_F n W L_1} \tag{1}$$

where *W* and $L_1$ are the width and length of graphene, respectively that absorb a light, $\eta_{abs}$ is the absorption efficiency, $P_{in}$ is the input power, $\tau_{el}$ is the electron-lattice scattering time, $E_F$ is the Fermi energy that scales with two dimensional carrier density *n* as $E_F = \hbar v_F (\pi n)^{1/2}$, $\hbar$ is the Planck constant and $v_F = 10^6$ m/s [4] is the Fermi velocity. As the total stored charge *n* in the graphene can be determined from the capacitor equation

$$en = C_{g/a}(V_g - V_t) \tag{2}$$

where $C_{g/a}$ is the gate capacitance per unit area defined as $C_g = \varepsilon_0 \varepsilon_d / t_d$, $\varepsilon_0$ is the vacuum permittivity, $\varepsilon_d$ is the relative dielectric constant and $t_d$ is gate dielectric thickness, $V_t$ is threshold voltage, *e* is the elementary charge. Taking into account the length *L* and width *W* of the photodetector, the eq. 2 can takes form

$$enWL = C_g(V_g - V_t) \tag{3}$$

where $C_g$ is the overall gate capacitance of the photodetector. Then

$$\frac{\Delta R}{R} = \frac{\tau_{el} e^{3/2}}{2\hbar v_F \pi^{1/2}} \frac{1}{[C_g(V_g - V_t)]^{3/2}} \eta_{abs} P_{in} \frac{W^{1/2} L^{3/2}}{L_1} \tag{4}$$

Here $L_1$ refers to the length of graphene photodetector that absorb a light and $L$ refers to the overall length of the photodetector. First term in eq. 4 depends on the graphene properties, the second term defines the total charge stored in the graphene under applied voltage, the third defines the absorbed power by the photodetector while the fourth term defines the strength of the bolometric photodetector. The gate capacitance in previous graphene-based photodetectors was calculated or measured at 6.5 fF [48], 11.5 fF [42], 20 fF [45], and 65 fF [48] that depends, as showed above, on the gate dielectric thickness and permittivity and photodetector dimensions. As observed from above, the thicker gate dielectric results in lower gate capacitance, thus the higher gate voltage will be required to achieve the same charge carriers in graphene and, in consequence, to set the maximum responsivity. The ratio of resistances and responsivities for different $V_b=\Delta V=V_g-V_t$ were presented in Fig. 2 for $\tau_{el}=1$ ps, $C_g=20$ fF, $L_1=10$ nm, $L=2$ μm, $W=40$ μm and $\eta_{abs}=40$ % (Supporting information). For shorter photodetector, $L=800$ nm, the gate capacitance was assumed at $C_g=50$ fF.

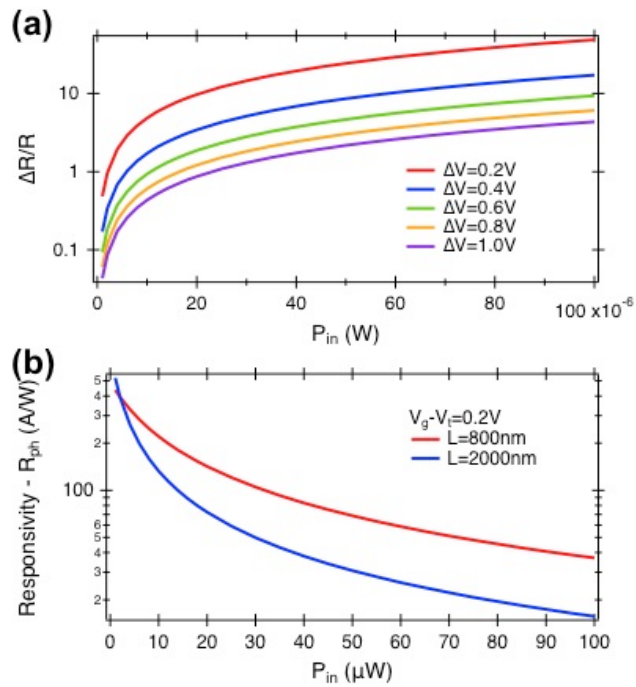

**Fig. 2.** (a) Ratio of resistances as a function of power absorbed by graphene sheet for different voltage increases $\Delta V=V_g-V_t$ for $W=40$ μm, $L=2$ μm and $L_1=10$ nm, and (b) responsivity as a function of absorbed power for different length of the photodetector $L$ and for $\Delta V=V_g-V_t=0.2$ V and $V_b=0.2$ V.

Under zero bias voltage, the PB photocurrent does not exist [33, 42]. However, under a bias voltage applied across the metallic contact, the change of graphene resistance can be detected by the change of the photocurrent flowing through the graphene sheet as [33, 54, 59]

$$I_{ph} = \Delta I = I_{off} - I_{on} = \frac{V_b}{R} - \frac{V_b}{R+\Delta R} = \frac{\Delta R}{R}\frac{V_b}{R+\Delta R} \qquad (5)$$

As observed, the photocurrent depends on the bias voltage $V_b$ and the change of graphene resistance that can be expresses by the ration of resistance $\Delta R/R$. Thus, the responsivity of photodetector is expressed by

$$R_{ph} = \frac{I_{ph}}{P_{in}} \tag{6}$$

Consequently, knowing the photocurrent of the photodetector and input power, the responsivity can be calculated. The calculations were performed for conductivity $\sigma_0$=0.4 mS, similar to Ref. [33]. For $L$=800 nm long and $W$=40 µm wide photodetector the resistance was calculated at $R$=50 Ω while the ratio of resistances for $V_b=V_g-V_t$=0.2 V, corresponding to $n$=7.8·$10^{14}$ m$^{-2}$ and $E_F$=0.2 eV for $C_g$=20 fF, was calculated at $\Delta R/R$=6.10 for input power of 50 µW and $\Delta R/R$=0.12 for lower input power of 1 µW. As observed from Fig. 2a (ratio of resistances), the PB photodetector works in inverse operation mode with the off-state in the dark (where the current signal is high) and the on-state with light incidence (where the current signal is low). Furthermore, to achieve a large on-off state a strong suppression of the current is highly desired with an applied optical signal. This observation is consistent with experimental work performed with the bow-tie photo-bolometric photodetector [33]. The current change between off and on state corresponds to a photocurrent.

When light is delivered to the photodetector with input power of 50 µW for $L$=800 nm long photodetector, that corresponds to the absorbed power of 20 µW, and voltage $V_b=V_g-V_t$=0.2 V the ratio of resistances was calculated at $\Delta R/R$=6.1 while the external responsivity at $R_{ph}$=67 A/W. For longer photodetector with $L$=2 µm, the resistance was calculated at $R$=125 Ω while the ratio of resistances at $\Delta R/R$=24. Thus, for $V_b=V_g-V_t$=0.2 V the responsivity was calculated at $R_{ph}$=31 A/W (Fig. 2b).

For lower input power of 10 µW, *i.e.*, absorbed power $P_{abs}$=4 µW and $L$=800 nm, $W$=40 µm photodetector, the external responsivity was calculated at $R_{ph}$=220 A/W while for $L$=2 µm it was calculated at $R_{ph}$=132 A/W. As observed from above, the low power operation is desired for the best performance of the photo-bolometric photodetector as it reduces power requirements and ensures enhanced responsivity that depends strongly on the ratio of resistances $\Delta R/R$ and bias voltage $V_b$ (eq. 5 and 6). Furthermore, as it was showed in our previous paper [34], the analysis of $\Delta R/R$ is strictly perturbative and it works very good for $\Delta R<R$, thus for lower input powers.

In the next section the responsivity of the proposed photodetector was compared with the state-of-the-art plasmonic PB photodetectors: bow-tie photodetector [33] and metal-insulator-metal (MIM) photodetector [42].

**Comparison with other graphene based plasmonic photo-bolometers**

Compared to the design proposed here (Fig. 3a), the PB plasmonic photodetector based on metal-insulator-metal (MIM) design was recently proposed, fabricated and characterized [42] (Fig. 3b) that provides a huge in-plane electric field on graphene as the result of a high field enhancement in a small gap. However, for a small gaps, the absorption losses in the metal arise, thus only 40 % of light can be absorbed by graphene for a gap width of only 15 nm. For wider gaps a graphene absorption decreases below 30 % for a gap width of 20 nm. However, it should be remembered that coupling efficiency from a silicon photonic waveguide to the MIM plasmonic waveguide was around 50 %. Thus, the overall absorption efficiency at graphene can be evaluated at 20 %. For an extremely small gap of 15 nm the responsivity was measured at $R_{ph}$=0.67 A/W for 8 µW input power and for wavelength of 1310 nm [42]. As expected, realization of such a device with an extremely small gap can be a very challenging.

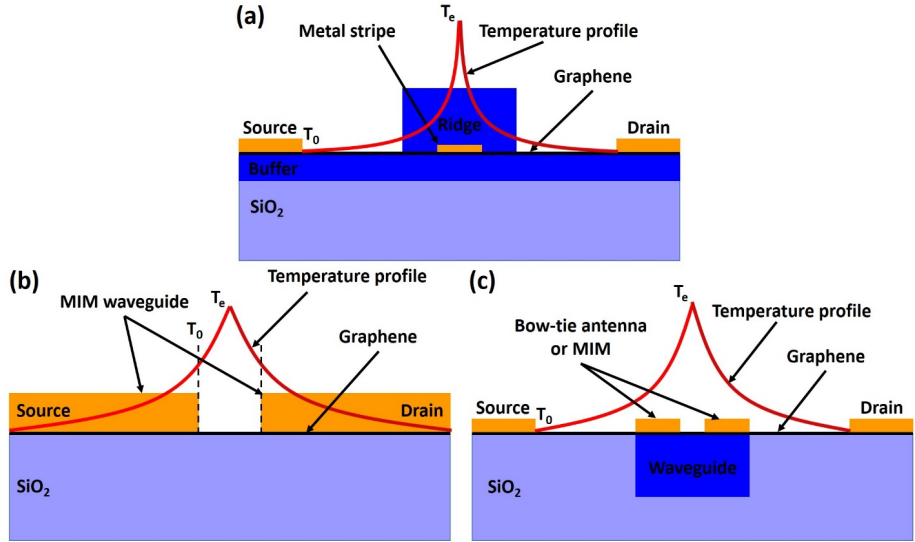

**Fig. 3.** (a) Cross-section of the photo-bolometric plasmonic photodetector in LR-DLSPP waveguide arrangement [34] with the corresponding plot of the electron temperature distribution between metal electrodes. (b, c) Cross-section of the MIM- [42] (b) and the bow-tie-based [33] (c) PB photodetectors with a corresponding temperature distribution between metal electrodes (red lines). (b) and (c) taken from the literature for comparison purposes [33, 42].

A similar arrangement was proposed, fabricated and characterized in Ref. [33] [Fig. 3c] where light from a Si photonic waveguide was coupled to pairs of bow-tie antennas placed on top of the Si waveguide. As a result, high field enhancement in the gap between the bow-tie antennas was observed which enhanced graphene absorption. In this arrangement graphene was placed directly below bow-tie antennas. In consequence, the absorption efficiency of 46 % in graphene for the five bow-tie antennas pairs was achieved for the gaps width ranging from 430 nm to 80 nm, *i.e.*, 255 nm in average. In this case, responsivity of 0.5 A/W was measured at wavelength of 1550 nm at input power of 80 μW. For a single bow-tie antenna pair, the absorption in graphene was calculated at ~23 %. The distance between source and drain (external electrodes/contacts) here was around 630 nm.

As showed in the main text of the manuscript, the relative resistance change, *ΔR/R*, is a key performance indicator of the bolometric photodetector. Based on the data provided in Ref. [33] and [42], we can compare those two PB photodetectors with our PB photodetector (Fig. 4).

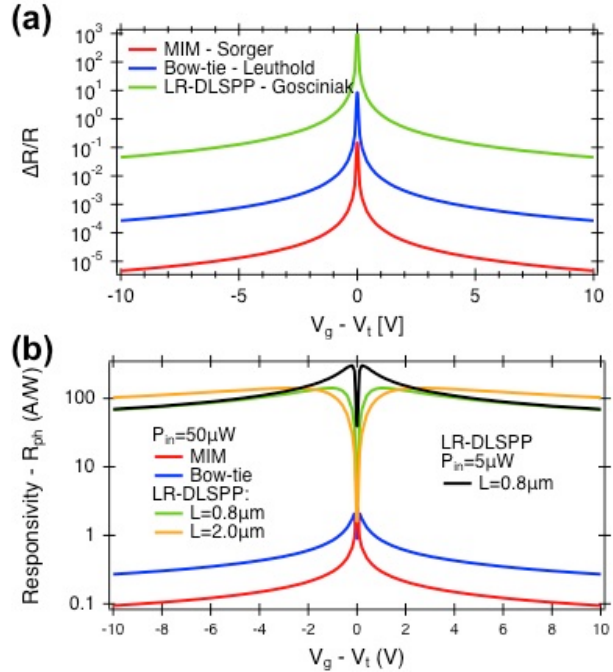

**Fig. 4.** (a) Comparison of waveguide-integrated plasmonic PB photodetectors for input power of 50 μW, gate capacitance $C_g$=20 fF and e-l scattering time of 1.0 ps. – (a) ratio of resistances $\Delta R/R$ and (b) responsivity $R_{ph}$.

As observed from Fig. 4a, the highest ratio of resistances is achieved for the LR-DLSPP arrangement [34] with $\Delta R/R$=24 ($L$=2 μm long photodetector) for $V_b=V_g-V_t$=0.2 V while for the MIM [42] and bow-tie [33] arrangements it was calculated at $\Delta R/R$=0.0017 and $\Delta R/R$=0.1, respectively for the same voltage $V_b=V_g-V_t$=0.2V. For MIM arrangement $L_1=L$=15 nm and $W$=5 μm while for bow-tie arrangement $L_1$=200 nm, $L$=630 nm and $W$=6 μm. The LR-DLSPP photodetector arrangement provides over 2 orders of magnitude higher ratio than bow-tie arrangement and over 4 orders of magnitude higher ratio than MIM arrangement. As observed from eq. 4, the ratio of resistances $\Delta R/R$ depends strongly on the ratio $L^{3/2}/L_1$ where $L_1$ refers to the length of graphene photodetector that absorb a light and $L$ refers to the overall length of the photodetector. For LR-DLSPP photodetector it was calculated at 8950 while for bow-tie and MIM arrangements at 80 and 3.9, respectively.

For bot-tie arrangement the resistance was measured at $R$=200 Ω [33]. Taking into account the ratio of resistances $\Delta R/R$=0.1 calculated for an input power of 50 μW, the external responsivity was calculated at $R_{ph}$=1.1 A/W for $V_b=V_g-V_t$=0.6 V (Fig. 4b). However, for $V_b=V_g-V_t$=1 V, the ratio of resistances was calculate at $\Delta R/R$=0.0085. Thus, the responsivity for an input power $P_{in}$=50 μW was calculated at $R_{ph}$=0.85 A/W. The maximum responsivity of 2.2 A/W was calculated for voltage $V_b=V_g-V_t$=0.1 V (Fig. 4b). On a contrary, for the MIM photodetector arrangement the resistance was calculated at $R$=10 Ω. For $V_b=V_g-V_t$=0.6 V the ratio of resistances was calculated at $\Delta R/R$=0.00032 what gives the external responsivity at $R_{ph}$=0.38 A/W. However, for $V_b=V_g-V_t$=0.2 V the external responsivity was calculated at $R_{ph}$=0.66 A/W what agree well with the experimental data $R_{ph}$=0.67 A/W [42].

The state-of-the-art plasmonic PB photodetectors proved to operate beyond 110 GHz under the room operation conditions [33]. As it has been previously reported, the operational speed of the PB photodetectors is not limited neither by the electrical RC delay time as it applies to the photo-voltaic (PV) effect [42], by transit time of carriers to the electrodes [20] nor by thermal speed of the device [48] but rather by relaxation time of hot carriers in graphene defined through the electron-lattice scattering time $\tau_{el}$ [33, 34, 42, 48]. Despite being much slower than electron-electron scattering time $\tau_{ee}$ that range from 10's of fs to 200 fs [60-62], it is still extremely fast ranging from 1 to 2 ps under the room temperature operation conditions [34, 63, 64]. It can be even further decreased by an increase of phonon-lattice temperature $T_L$. It is in a cost of sensitivity as a reduction in transport current decreases the $\Delta R/R$ while an increase in electron temperature $T_e$ is equivalent to an increase of carrier density and leads to an increase in transport current [34]. Through increases of the lattice temperature $T_L$, the more efficient cooling pathway for hot electrons is achieve as additional phonons become available for heat dissipation [11, 54, 61]. As it has been previously reported [34], the lattice temperature can be increased either through a Joule heating, *i.e.*, applied electrical power [11, 61] or through highly confined plasmonic energy [54]. Thus, the compact devices can provide additional heat dissipation channel through the increases of $T_L$ but, simultaneously, reducing a PB effect [42]. It should be however noted that the efficiency of electron heating is independent of lattice temperature and depends only on the in-plane component of the electric field coupled to the graphene [34, 61]. Recent studies showed that even rise the electron temperature $T_e$ to 1000's of K does not substantially move the lattice temperature $T_L$ away from the ambient temperature $T_0$ [65].

**Evaluation of responsivity from electron temperature rise in a graphene photodetector**

As it was mentioned in our previous paper [34] and repeated here, the proposed analysis of the PB effect is perturbative and thus approximate, so mostly is valid for the $\Delta R<R$. For large change of resistance, *i.e.*, $\Delta R>R$, the electron temperature should be estimated directly.

The general photocurrent in a bolometric photodetector of channel width $W$ and length $L$ is determined by electron temperature $\Delta T_e$ and lattice temperature $\Delta T_l$ as [48]

$$I_{ph} \approx \frac{W}{L}\left(\frac{\partial \sigma}{\partial T_e}\bigg|_{T_e=T_l=T_s}\Delta T_e + \frac{\partial \sigma}{\partial T_l}\bigg|_{T_e=T_l=T_s}\Delta T_l\right)(V_b) \quad (7)$$

where $\partial\sigma/\partial T_e$ and $\partial\sigma/\partial T_l$ are the rates of conductivity variation against changes in electron and phonon temperatures and $T_s$ is an initial temperature at equilibrium without external illumination. As observed, the photocurrent $I_{ph}$ depends on the electron temperature $\Delta T_e$ and phonon temperature $\Delta T_l$ variations. Thus, it is required to evaluate the electron temperature rise in the photodetector under absorbed power. The absorbed power has two main contributions – one related with a light absorption and the second with electrical Joule heating. The Joule heating power is produced by the external bias voltage $V_b$ and is defined as $P_{abs}=V_b^2/R$ [48].

The carrier concentrations in graphene and thus the electron temperature of hot electrons in graphene $T_e$ in a function of absorbed power is governed by the heat transfer equation [66, 67]

$$\kappa_e \nabla^2 T_e - g_{e-l}(T_e - T_0) + P^* = 0 \quad (8)$$

where $\kappa_e$ represents the graphene thermal conductivity, $g_{e-l}$ is the thermal coupling between electrons and lattice ($g_{e-l}=\gamma C_e$), $\gamma$ is the electron-lattice (e-l) cooling rate, $C_e$ the electron heat capacity, $T_0$ is the temperature of the substrate and $P^*$ is the power per unit area absorbed by the electrons (Supporting Information 1). All those parameters govern the heat dissipation in graphene and hence determine the electron temperature rise and distribution in graphene.

The electronic thermal conductivity $\kappa_e$ is obtained from the Wiedemann-Franz relation that relates the thermal and electrical conductivity of metals and is expressed by [66, 67]

$$\kappa_e = \frac{\pi^2 k_B^2 T_e}{3e^2}\sigma \tag{9}$$

where $e$ is the electron charge, $T_e$ is the operating temperature and $\sigma$ is the electrical conductivity of graphene. The photon energy provided to the graphene quickly thermalize among the graphene electrons due to the fast e-e interaction time, which occurs on a femtoseconds time scale [60-63]. As a result, a quasi-equilibrium is established in tens of femtoseconds [60, 63].

The initial temperature rise in graphene is determined using the electronic heat capacity of monolayer graphene [68, 69]

$$C_e = A\gamma T_e \tag{10}$$

where $A$ is the area of the graphene sheet and

$$\gamma = \left(4\pi^{5/2} k_B^2 \sqrt{n}\right)/(3h v_F) \tag{11}$$

is the Sommerfeld coefficient where $k_B$ is the Boltzmann constant, $h$ is the Planck constant, $v_F=10^6$ m/s is the graphene Fermi velocity and $n$ is the charge-carrier density. The calculated electronic thermal conductivity $\kappa_e$ and electronic heat capacity $C_e$ as a function of Fermi energy for different electron temperatures $T_e$ were presented in Fig. S1. As observed, the electronic thermal conductivity is very sensitive to the Fermi energy and grows fast for higher carrier concentrations, Fermi energy. Furthermore, it is much higher for higher electronic temperatures in a graphene (Fig. S1a).

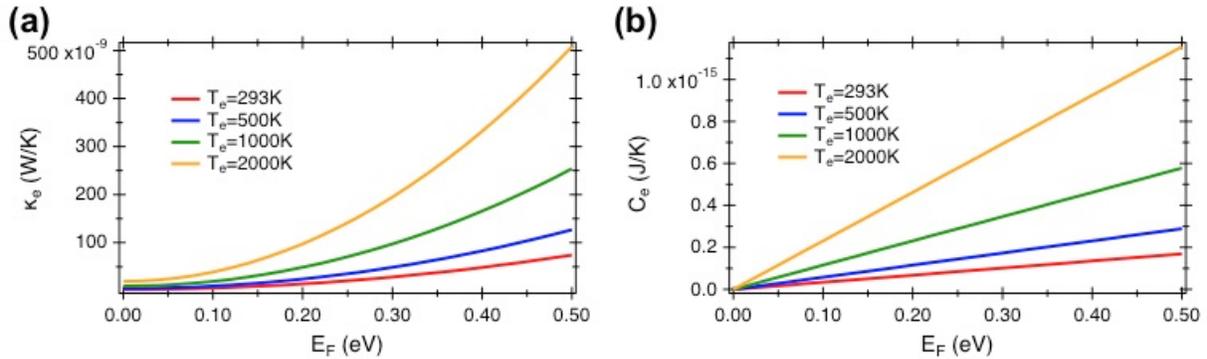

**Fig. 5.** (a) The electron thermal conductivity and (b) electronic heat capacity as a function of Fermi energy, carrier concentration, at electron temperatures ranging from $T_e$=293 K to $T_e$=2000 K. The electron thermal conductivity calculated for a charge neutrality width $\Delta$=100 meV.

In metals, the specific heat is stored by the lattice vibrations (phonons) and the free conduction electrons. In graphene, phonons dominate the specific heat of graphene at all temperatures above 1 K, and the phonon specific heat increases with temperature. At very high temperatures, the lattice heat capacity is nearly constant at $C_p$=~5·10$^{-14}$ J/K while the electronic heat capacity $C_e$ range from ~1.2·10$^{-15}$ to 2·10$^{-18}$ J/K (Fig. 5). Consequently, the lattice heat capacity is ~1-3 orders higher than the electronic heat capacity [65]. Due to it, the phonon system can be treated as an ideal thermal bath with $T_0$ staying constant while the absorbed power can give rise to an electron temperature $T_e$ that can be dramatically higher than the lattice temperature $T_0$ [64]. Furthermore, $C_e$ can be tuned using a gate voltage with a minimum at the charge neutrality point. Because of much lower $C_e$ compared to $C_p$, the power absorbed by graphene rise the steady state electron temperature $T_e$ of the graphene above the lattice temperature $T_0$.

The analytical solution to the heat equation of the system along a photodetector length L is [64]

$$\Delta T_e(y) = T_e(y) - T_0 = \frac{sinh((0.5 \cdot L - |y|)/\xi)}{cosh((0.5 \cdot L)/\xi)} \frac{P^*}{2} \frac{\xi^2}{\kappa_e} \quad (12)$$

where L is the length of the photodetector and ξ is the e-l cooling length for hot carrier propagation in the photodetector. The e-l cooling length ξ is a combination of $\kappa_e$ and $g_{el}$ through $\xi=(\kappa_e/g_{el})^{1/2}$.

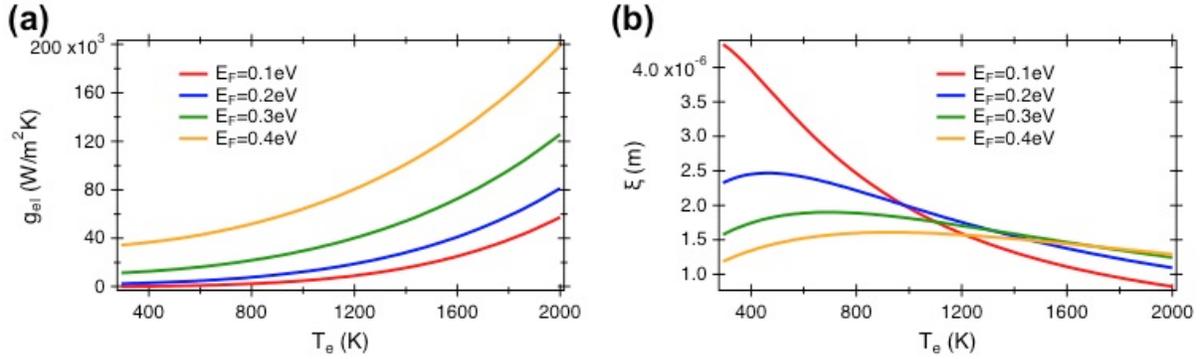

**Fig. 6.** Temperature dependent (a) e-ph coupling parameter $g_{el}$ and (b) e-l cooling length ξ for different Fermi energies.

The electron temperature dependent $g_{el}$ and ξ for different Fermi energies, carrier concentrations, were showed on Fig. 6. As observed, the e-l cooling length ξ strongly depends on the Fermi energy, carrier concentration. For Fermi energy $E_F$=0.1 eV, the cooling length reaches ξ=4.4 μm and drops to ξ=1.2 μm for $E_F$=0.4 eV at $T_e$=300 K. As the cooling length decreases, the hot carriers strongly thermalize with the lattice leading to reduction of electron temperature.

For optimal operation of the photo-bolometric photodetector a distance between photogenerated carriers in graphene and external electrodes should exceed an electron mean free path $l_{MFP}$. Under such circumstances, the electron temperature $T_e$ will be at maximum while the acoustic phonons and the electrons in the contacts will remain at the bath temperature $T_0$, i.e., preferably close to the room temperature. As shown (Fig. 4), the proposed photodetector arrangement enables a realization of photodetectors with a distance between electrodes that exceeds the electron mean free path that was calculated at $l_{MFP}$=38 nm. As such, the ballistic heat transfer through the contacts is reduced. In Fig. 7 calculations were performed for conductivity $\sigma_0$=0.2 mS.

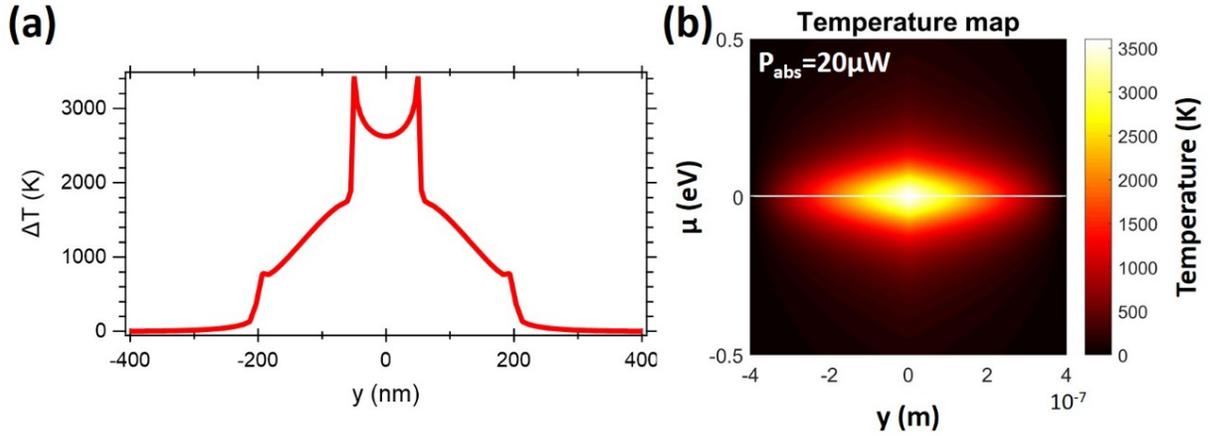

**Fig. 7.** (a) Temperature distribution between Metal Pad 1 and Metal Pad 2 (Fig.1) for $E_F$=0 eV and (b) temperature map along metal contacts and for different chemical potentials $\mu$. Here, conductivity was taken at $\sigma_0$=0.2 mS and length of the photodetector at $L$=1000 nm.

Furthermore, in Fig. 8 the electron temperature rise in the LR-DLSPP photodetector (Fig. 8a) was compared with other plasmonic bolometric photodetectors – MIM (Fig. 8b) and bow-tie (Fig. 8c). As observed, the highest temperature rise is observed for LR-DLSPP arrangement (Fig. 8a) where electron temperature exceeds $\Delta T_e$=4000 K in the middle of the photodetector. In comparison, for MIM arrangement it slightly exceeds $\Delta T_e$=170 K (Fig. 8b) while for bow-tie arrangement it reaches $\Delta T_e$=960 K (Fig. 8c) under the same input power. It has been recently reported that electron temperature in graphene can increase up to 5000 K under illumination [65].

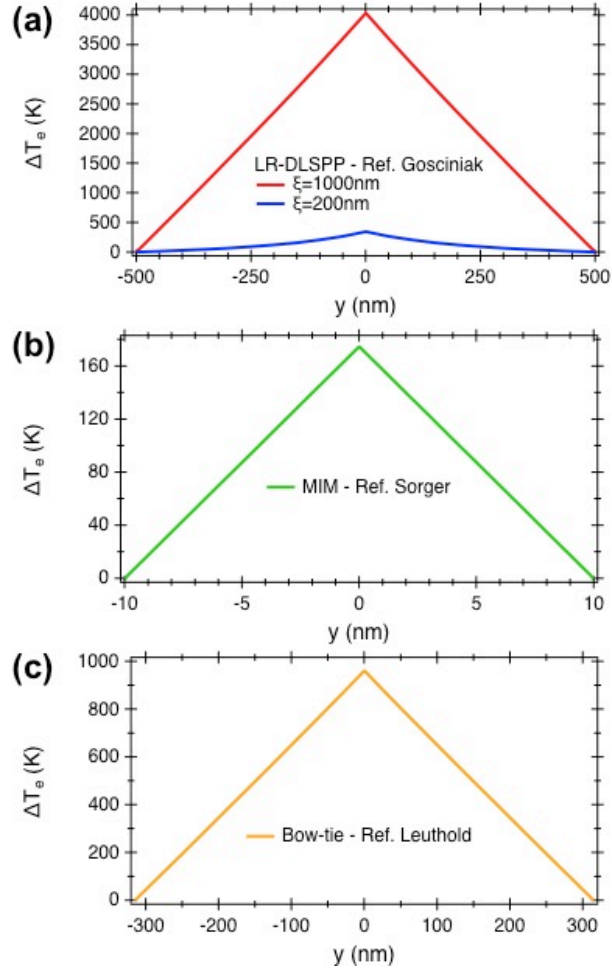

**Fig. 8.** Temperature distribution between external electrodes for (a) LR-DLSPP [34], (b) MIM [42] and (c) bow-tie photodetector [33] arrangements. Here, conductivity was taken at $\sigma_0$=0.4 mS, input power at $P_{in}$=50 μW and cooling length at $\xi$=1000 nm while for (a) an additional cooling length $\xi$=200 nm was considered. The length of photodetector was taken at (a) $L$=1000nm, (b) $L$=20 nm and (c) $L$=630 nm.

From calculated electron temperature rise $\Delta T_e$ in a graphene it is possible to calculate the photocurrent of the bolometric photodetector using eq. 7. As observed, the higher electron temperature increases in graphene the higher photocurrent $I_{ph}$. For $W$=40 μm, $L$=1 μm, $\Delta T_e$=4000 K as showed in Fig. 8a and $\partial\sigma/\partial T_e$=0.5 μS/K as provided in Ref. [48], the photocurrent was calculated at $I_{ph}$=16 mA. The temperature rise in the graphene sheet in Fig. 8a was calculated for the input power of 50 μW, thus the responsivity was calculated at $R_{ph}$=180 A/W. In contrast, the responsivity calculated from the ratio of resistances $\Delta R/R$ for the same parameters was calculated at $R_{ph}$=130 A/W for $V_b$=$V_g$-$V_t$=0.6 V (Fig. 4b). For bow-tie photodetector arrangement [33], the photocurrent was calculated at $I_{ph}$=0.9 mA what gives a responsivity $R_{ph}$=9 A/W. In comparison, from $\Delta R/R$ calculation the responsivity was calculated at $R_{ph}$=0.8-1.1 A/W. However, large inconsistency appears in case of MIM photodetector arrangement [42] where photocurrent was calculated at $I_{ph}$=1.75 mA and responsivity at $R_{ph}$=35 A/W. It is much higher value than calculated from $\Delta R/R$ where responsivity was estimated at $R_{ph}$=0.66 A/W. This inconsistency results from the extremely small gap of 15 nm between electrodes that give rise to additional heat flow channel to the external electrodes. When compared with the experimental data we see that responsivity for bow-tie photodetector was measured at $R_{ph}$=0.5 A/W while for MIM photodetector at $R_{ph}$=0.67 A/W. As observed, the responsivity agree pretty well with the calculations performed based on the $\Delta R/R$ – $R_{ph}$=0.80 A/W and $R_{ph}$=0.66 A/W for bow-tie and MIM photodetectors, respectively.

For a carrier mobility $\mu_m$=1900 cm$^2$/(V·s), as measured at Ref. [33] and that corresponds to conductivity $\sigma$=0.4 mS at the charge neutrality point and relaxation time of 1 ps, a diffusion length $L_D=(V_t\mu\tau)^{1/2}$ of 140 nm was calculated for $V_t$=0.1 V. It is substantially less than the length of the photodetector in LR-DLSPP (L=800-2000 nm) and bow-tie (L=630 nm) arrangements. Consequently, the $T_0$ under those circumstances stay at room temperature of 293 K. However, for more compact photodetector presented in Ref. [42], the diffusion length highly exceed the length of the photodetector of 15 nm, thus the temperature $T_0$ at contact highly increases. As a result, the temperature increases $\Delta T_e$ between the center of graphene channel and contacts is low (Fig. 8b), what highly reduced the bolometric effect. For a very compact photodetectors as in Ref. [42], a distance between electrodes is very close to the average separation between electrons defined as $D=(1/n)^{1/2}$ where $n$ is the carrier concentration in graphene [70]. For a typical $n$=1·10$^{12}$ cm$^{-2}$, $D$=10 nm and decreases for higher carrier concentrations. As a result, the screening of hot carriers through the external electrodes takes place what further reduces the electron temperature in graphene. As stated in Ref. [42], a reduction on slot width give rise to an increased ballistic transport conditions [71].

**Conclusion**

Here, a waveguide-integrated plasmonic graphene photodetector operating based on the hot carriers photo-bolometric effect was evaluated and compared with other state-of-the-art bolometric photodetectors. A comparison was performed based on a theory of the bolometric effect originating from the band nonparabolicity of graphene showing a responsivity exceeding 200 A/W for LR-DLSPP-based photodetector. Based on the same theory, a responsivity of bow-tie-based photodetector was calculated at 0.8 A/W while for MIM-based photodetector is was calculated at 0.66 A/W. The results were compared with a standard theory where a photocurrent, and in consequence, a responsivity are determined by the electron temperature increases in a graphene channel. It shows a good agreement with our developed theory for photodetectors with a length exceeding a diffusion length. And what is even more important, it shows very good fit with the experimental data for MIM and bow-tie photodetectors.

**Author information**

**Affiliations**

Independent Researcher, 90-132 Lodz, Poland

Jacek Gosciniak

John Hopkins University, Baltimore, MD 21218, USA

Jacob Khurgin

**Contributions**

J.G. performed calculations and FEM and FDTD simulations and wrote the article. J.G and J.B.K analyzed and discussed the results. Both authors reviewed the article.

**Corresponding author**

Correspondence to Jacek Gosciniak (jeckug10@yahoo.com.sg)

**Supporting information**

The power absorbed by the graphene photodetector $P_{abs}$ is related to the input power $P_{in}$ as $P_{abs}=\eta_{abs}\eta_c P_{in}$, where $\eta_c$ and $\eta_{abs}$ are the coupling and absorption efficiencies, respectively. Here $L_1$ is the length of the in-plane electric field interacting with graphene. The coupling efficiency in this type of plasmonic waveguide can exceed 90 %, with absorption efficiency exceeding 40 % for 40 µm-long and up to 63 % for 100 µm-long photodetectors, respectively (Fig. S1). Simultaneously, as showed in Fig. 1b, the power is absorbed by less than 10 nm-wide graphene sheet. As a result, the power absorbed by the graphene is extremely high, enhancing the electron temperature in the graphene (Fig. 8).

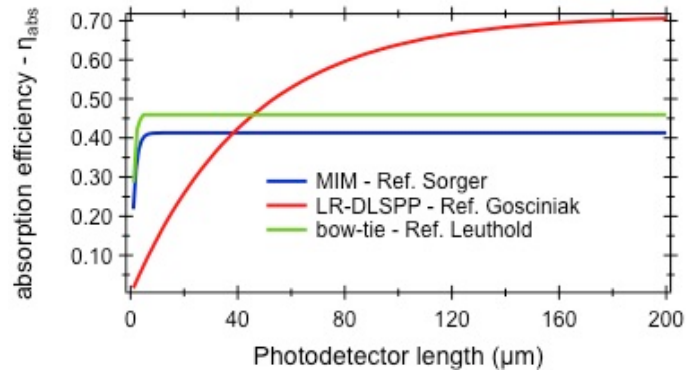

**Fig. S1.** (a) Comparison of the light absorption efficiencies in graphene for LR-DLSPP-based [34], MIM-based [42] and bow-tie-based photodetectors [33].

The results were compared with the state-of-the-art MIM [Sorger] and bow-tie [Leuthold] photodetectors showing absorption efficiency of 41 % and 46 % for MIM and bow-tie arrangements, respectively [Fig. S1]. As observed, it is lower than for LR-DLSPP arrangements, however, it is achieved for only 5 µm long waveguides.